\documentclass[preprintnumbers, floatfix, letterpaper, superscriptaddress,nofootinbib, twocolumn]{revtex4}
\usepackage[dvipdfmx]{graphicx}
\usepackage{microtype}
\usepackage{amsmath}
\usepackage{amssymb}
\usepackage{subfigure}
\usepackage{hyperref}
\usepackage{float,color} 
\usepackage{url}
\usepackage{xcolor}
\usepackage{color}
\usepackage{mathrsfs}
\usepackage{amsfonts}
\usepackage{latexsym}
\usepackage{ragged2e}
\usepackage{epsfig}
\usepackage{textcomp}
\usepackage{phaistos}

\makeatletter
\renewcommand\@makefnmark{\hbox{\@textsuperscript{\normalfont\color{purple}\@thefnmark}}}
\renewcommand\@makefntext[1]{%
  \parindent 1em\noindent
            \hb@xt@1.8em{%
                \hss\@textsuperscript{\normalfont\@thefnmark}}#1}
\makeatother

\usepackage{caption}
\DeclareCaptionJustification{justified}{\leftskip=0pt \rightskip=0pt \parfillskip=0pt plus 1fil}
\def\beq{\begin{equation}}
\def\eeq{\end{equation}}

\def\mathbb{\Bbb}

\newcommand{\alf}{Alfv\'en\ }
\newcommand{\alfv}{Alfv\'enic\ }

\newcommand{\vp}{\varphi}

\newcommand{\pa}{\partial}
\newcommand{\df}{\dfrac}

\definecolor{vividviolet}{rgb}{0.62, 0.0, 1.0}
\definecolor{amaranth}{rgb}{0.9, 0.17, 0.31}
\definecolor{palatinateblue}{rgb}{0.15, 0.23, 0.89}
\definecolor{brightpink}{rgb}{1.0, 0.0, 0.5}
\definecolor{cornflowerblue}{rgb}{0.39, 0.58, 0.93}
\definecolor{deepcarminepink}{rgb}{0.94, 0.19, 0.22}
\definecolor{radicalred}{rgb}{1.0, 0.21, 0.37}
\colorlet{RED}{red}

\hypersetup{ linktoc=all,
    colorlinks, linkcolor={palatinateblue},
    citecolor={brightpink}, urlcolor={amaranth}
}

\graphicspath{{Images/}}

\graphicspath{{Images/}}

\makeatletter
\def\@fnsymbol#1{\ensuremath{\ifcase#1
\or $\textleaf$ \or $\PHplaneTree$ \or $\PHrosette$ \or $\PHvine$
\else\@ctrerr\fi}}%

\makeatother

\def\sideremark#1{\ifvmode\leavevmode\fi\vadjust{\vbox to0pt{\vss
 \hbox to 0pt{\hskip\hsize\hskip1em
 \vbox{\hsize1.5cm\tiny\raggedright\pretolerance10000
 \noindent #1\hfill}\hss}\vbox to8pt{\vfil}\vss}}}%
\begin{document} 
\title{\alfv superradiance for a monopole magnetosphere around a Kerr black hole}
\author{Sousuke \surname{Noda}}
\email{snoda@cc.miyakonojo-nct.ac.jp}
\affiliation{National Institute of Technology, Miyakonojo College, Miyakonojo 885-8567, Japan}
\affiliation{Center for Gravitation and Cosmology, College of Physical Science and Technology, Yangzhou University, \\180 Siwangting Road, Yangzhou City, Jiangsu Province  225002, China}

\author{Yasusada \surname{Nambu}}
\email{nambu@gravity.phys.nagoya-u.ac.jp}
\affiliation{Department of Physics, Graduate School of Science, Nagoya University, \\
  Chikusa, Nagoya 464-8602, Japan}
  
  \author{Masaaki \surname{Takahashi}}
 \email{mtakahas@auecc.aichi-edu.ac.jp}
\affiliation{Department of Physics and Astronomy, Aichi University of
  Education, Kariya, Aichi 448-8542, Japan}
  
  \author{Takuma \surname{Tsukamoto}}
\affiliation{Department of Physics, Graduate School of Science, Nagoya University, \\
  Chikusa, Nagoya 464-8602, Japan}

\begin{abstract}
Herein, we explore superradiance for Alfv\'en waves (Alfv\'enic superradiance) in an axisymmetric 
rotating magnetosphere of a Kerr black hole within the force-free approximation. 
On the equatorial plane of the Kerr spacetime, the \alf wave equation is reduced to a one-dimensional Schr\"{o}dinger-type equation by separating variables of the wave function and introducing a tortoise coordinate mapping the inner and outer light surfaces to $-\infty$ and $+\infty$, respectively, and we investigate a wave scattering problem 
for \alf waves. 
An analysis of the asymptotic solutions of the wave equation and conservation of the Wronskian provides the superradiant condition 
for \alf waves, and it is shown that the condition coincides with that for the Blandford-Znajek process. 
This indicates that when \alfv superradiance occurs, the Blandford-Znajek process also occurs in the force-free magnetosphere.
Then, we evaluate the reflection rate of \alf waves numerically and confirm that 
\alfv superradiance is indeed possible in the Kerr spacetime. 
Moreover, we will discuss a resonant scattering of \alf waves, which is 
related to a ``quasinormal mode'' of the magnetosphere.
\end{abstract}


\maketitle

\section{Introduction}
The mechanism for extracting the rotational energy from a black hole has been discussed as an energy source for high-energy astronomical phenomena such as relativistic jets in active galactic nuclei, compact objects, and gamma-ray bursts. 
The Blandford-Znajek (BZ) process \cite{Blandford1977} is one of the most promising 
candidates to describe this mechanism, which is driven by rotating black hole magnetosphere. 
The original BZ process \cite{Blandford1977} was discussed with focus on Kerr spacetime and force-free magnetospshere, for which the plasma inertia is ignored due to the strong electromagnetic fields. Then, they discovered that if the angular velocity of the rotating black hole $\Omega_\text{H}$ exceeds that of magnetic field lines $\Omega_F$: 
 \beq
 0<\Omega_F <\Omega_\text{H}, 
 \label{eq:BZ_cond}
 \eeq
 the rotational energy of the black hole is transported outward in the form of the Poynting flux, 
 which is caused by the magnetic torque acting on magnetic field lines due to the dragging effect 
 of the rotating black hole. 
After the pioneering work by Blandford and Znajek in 1977 \cite{Blandford1977}, 
several supportive works have been conducted based on analytical and numerical computations not only 
for force-free manetosphere, but also for 
magnetohydrodynamic case, for example \cite{Toma2014,Toma2016,Jacobson2019,Kinoshita2018,Komissarov2004,Komissarov2005,
Koide2006,McKinney2006,Ruiz2012,Koide2014,Koide2018}.

Superradiance \cite{Zeldovich1971,Zeldovich1972,Starobinsky1973,Starobinsky1974,Brito2015} 
is also an energy extraction mechanism, which is often described as a wave version of the Penrose process \cite{Penrose,Wagh1985,Dadhich2012}.
The condition for superradiance (superradiant condition) is given by 
\beq
0<\df{\omega}{m}<\Omega_\text{H},
\label{eq:superrad_cond}
\eeq
where $\omega$ is the frequency of a wave and $m$ is the azimuthal quantum number.
As various wave phenomena will occur in the magnetosphere, the effect of energy extraction 
by waves should also be considered.

Although, in general, the BZ process and superradiance are considered as different mechanisms, conditions \eqref{eq:BZ_cond} and \eqref{eq:superrad_cond} appear similar regarding the ratio 
$\omega/m$ as the angular velocity of a wave pattern. 
There must be various wave modes in a black hole magnetosphere, hence investigation of the relationship between the BZ process 
and superradiance is important not only for clarifying their mathematical relation, but also for understanding the essence 
of the BZ process. Indeed, there are several works on superradiant scattering of waves in black hole magnetospheres: superradiance 
for the fast magnetosonic wave \cite{Uchida1997c,Putten1999} and energy extraction via scalar clouds as a proxy for the 
force-free magnetosphere \cite{Wilson-Gerow2016}, but the relationship between superradiance and the BZ process had not been 
clarified until our previous work \cite{Noda2020}.
In our previous work \cite{Noda2020}, we investigated the relationship between the BZ process and 
superradiance by discussing the superradiant scattering of \alf waves (\alfv superradiance) for a force-free magnetosphere 
in Ba\~nados-Teitelboim-Zanelli (BTZ) black string spacetime \cite{Jacobson2019}, and suggested that the BZ process is the zero 
mode of \alfv superradiance. The BTZ black string spacetime is asymptotically anti-de Sitter spacetime and its horizon geometry is cylinder, hence, it is not an astrophysical black hole. As black hole candidates 
observed so far are well-explained with Kerr black hole, it is important to check whether or not the \alfv superradiance is possible for a force-free 
magnetosphere around a Kerr black hole.

One of the differences between the magnetospheres in the BTZ string spacetime and the
Kerr spacetime is the existence of the outer light surface. For the Kerr spacetime case there is an outer light surface 
that provides outgoing one-way boundary condition to \alf waves.
If we solve the wave equation with the outgoing boundary condition at the outer light surface, which is similar to the 
computation of black hole quasinormal modes, it is possible to discuss the stability of the magnetosphere for the perturbation 
associated with \alf waves.

  In this paper, we investigate the possibility of the \alfv superradiance in the Kerr spacetime. 
  To achieve this, we solve the equation for force-free black hole magnetosphere in the Kerr spacetime to obtain a 
  background magnetosphere. 
  Then, we apply a perturbation to it and discuss the wave propagation in the background black hole magnetosphere. However, the global magnetosphere around a Kerr black hole is difficult to 
 obtain as we need to solve the Grad-Shafranov equation \cite{Grad,Shafranov,Blandford1977} in the Kerr spacetime. 
  Hence, our computation will be restricted to the electromagnetic 
  field in the vicinity of the equatorial plane of the Kerr spacetime. 
  Moreover, the force-free magnetosphere is assumed to be symmetric about 
  the equatorial plane, axisymmetric, and stationary. 
  Then, applying an appropriate perturbation to the background magnetosphere, the wave equation 
 for \alf waves will be derived.

    This paper is organized as follows. In section II, we review the 
    force-free electromagnetic field and obtain the background magnetosphere around the equatorial plane of the 
    Kerr spacetime and confirm whether the BZ process is possible for the background magnetosphere 
    solution. In section III, the \alf wave equation will be derived 
    by giving a perturbation to the background magnetosphere. 
    Then, we rewrite the wave equation in the form of the  
    Schr\"{o}dinger-type equation to clarify the propagation and scattering problem of \alf waves with 
    the effective potential. 
    Section IV presents the discussion of \alfv superradiance and the derivation of 
    the superradiant condition, and the reflection rates of \alf waves are evaluated with a numerical 
    calculation. 
    Furthermore, we discuss a resonant scattering of \alf waves, which is related to 
    a ``quasinormal mode'' of the background magnetosphere.
     The conclusion is provided in section V. We use the CGS units
      in electromagnetism and $c=G=1$ throughout this paper.

\section{Force-free electromagnetic field in the Kerr spacetime}
First, we briefly review the force-free approximation of the 
plasma-electromagnetic field system in a curved spacetime. Applying it to the 
Maxwell equation in the Kerr spacetime, we obtain a configuration of force-free 
electromagnetic field in the vicinity of the equatorial plane of the Kerr spacetime. Then, we discuss 
the BZ process for the background magnetosphere solution.
\subsection{Force-free approximation}
The basic equations are 
Maxwell's equation with electric 4-current $j^{\mu}$,
    \begin{equation}
        \nabla_{\alpha}F^{\mu\alpha}=4\pi j^{\mu},\ \ 
        \nabla_{[\mu}F_{\nu\lambda]}=0,
        \label{eq:Maxwell}
    \end{equation} 
and conservation of the energy-momentum tensor,  
    \begin{equation}
      \nabla_{\nu}(T^{\mu\nu}_\text{plasma}+T_\text{em}^{\mu\nu})=0,
        \label{eq:EM_tensor}
    \end{equation} 
where $T^{\mu\nu}_\text{plasma}$ is the energy 
momentum tensor of plasma and $T_\text{em}^{\mu\nu}$ is that of electromagnetic 
field. If the electromagnetic fields are so strong that the inertia of plasmas can be 
ignored, the above conservation law becomes $\nabla_{\nu}T_\text{em}^{\mu\nu} \simeq 0$. 
This is the force-free approximation. Hereafter, we simply denote $T_\text{em}^{\mu\nu}$ 
as $T^{\mu\nu}$, and it is given by
\begin{equation}
   T_{\mu\nu}=
   F_{\mu\alpha}F_{\nu}^{\ \alpha}-\frac{1}{4}F_{\alpha\beta}F^{\alpha\beta}g_{\mu\nu}.
   \label{eq:energy-momentum}
 \end{equation} 
Using the force-free approximation, it can be shown that 
$\nabla_{\nu}T_{\ \mu}^{\nu}=-4\pi F_{\mu\nu}j^{\nu} \simeq 0$. 
Therefore, the Maxwell equation under the force-free approximation is
 \begin{equation}
    F_{\mu\nu}\nabla_{\alpha}F^{\nu\alpha}=0,\quad
    \nabla_{[\mu}F_{\nu\lambda]}=0.
    \label{eq:basic}
 \end{equation} 

 For an observer of which 4-velocity is given by $u^\mu$, the electric field $E^{\mu}$ 
  and the magnetic filed $B^{\mu}$ are defined as $E^{\mu}=F^{\mu\nu}u_{\nu}$ 
  and $B^{\mu}=-{}^*F^{\mu \nu}u_{\nu}$, respectively.  The field strength 
  $F_{\mu\nu}$ is assumed to be magnetically dominated, as 
\begin{equation}
  F_{\mu \nu }F^{\mu \nu}
  =2(B^{\mu}B_{\mu}-E^{\mu}E_{\mu})>0.
 \label{magdominant}
\end{equation} 
This condition ensures the existence of a timelike observer who only sees the magnetic 
field. The field strength satisfying Eq. \eqref{eq:basic} can be represented with the 
Euler potentials $\phi_1$ and $\phi_2$ \cite{Uchida1997,Uchida1997a,Uchida1997b,Uchida1997c,Gralla2014} 
as
\begin{equation}
  F_{\mu\nu}
  =\partial_{\mu}\phi_{1}\partial_{\nu}\phi_{2}-\partial_{\mu}\phi_{2}\partial_{\nu}\phi_{1},
  \label{eq:2form}
 \end{equation}
and the Maxwell equation with the force-free approximation yields the following nonlinear 
equations for the Euler potentials:
 \begin{equation}
  \partial_{\mu}\phi_i \partial_{\nu}\left[\sqrt{-g}\left(\partial^{\mu}\phi_1\partial^{\nu}\phi_2-\partial^{\nu}\phi_1\partial^{\mu}\phi_2\right)\right]=0,\ \ (i=1, 2).
        \label{eq:basicEuler}
 \end{equation} 
By solving these two equations, we obtain the Euler potentials and the field strength. In the next subsection, 
we present a solution around the equatorial plane of the Kerr spacetime with arbitrary values of the spin parameter.
\subsection{Background force-free magnetosphere}
As a background magnetosphere to investigate the propagation of \alf waves, we obtain a force-free
magnetosphere solution with monopole-like magnetic field lines around the equatorial plane of a 
Kerr black hole by solving Eq.~\eqref{eq:basicEuler} with the fixed Kerr metric. 
First, let us introduce the Boyer-Lindquist coordinates 
$(t,r,\theta,\varphi)$ of the Kerr spacetime
\begin{align}
\notag    g=&
	-\left(1-\df{2M r}{\Sigma}\right)dt^2-\df{4aMr \sin^2{\theta}}{\Sigma}dt d\vp\\
	&+\dfrac{\Sigma}{\Delta}dr^2+\Sigma d\theta^2+\df{A \sin^2{\theta}}{\Sigma}d\vp^2,
         \label{eq:kerr}
 \end{align} 
 where $\Delta=r^2-2Mr+a^2$\ , \ $\Sigma=r^2+a^2\cos^2{\theta}$\ ,\
   $A=(r^2+a^2)^2-\Delta a^2 \sin^2{\theta}$, and the constants $M$ and $a$ are the mass 
   and angular momentum per unit mass of the Kerr black hole, respectively. 
   The outer horizon radius $r_{\text{H}}$ is given as the larger root of $\Delta=0$. 
  The dragging of spacetime is represented by the angular velocity of the zero angular momentum observer
  \beq
  \Omega(r):=-\df{g_{t\vp}}{g_{\vp\vp}},
  \eeq
and the value of this function at the outer horizon is $\Omega _\text{H} := a/(2Mr_\text{H})$. 
The Kerr spacetime has two Killing vectors, 
${\xi}_{(t)}={\bf{\pa}}_t$ and ${\xi}_{(\varphi)}={\bf{\pa}}_\varphi$. The region where the timelike Killing vector 
becomes spacelike is called the ergoregion. 

The solution of a monopole-type magnetic field for a force-free magnetosphere 
around the equatorial plane is given as
 \begin{equation}
       \phi_{1}=q \cos{\theta},\ \phi_{2}=\varphi-\Omega_F t  + J_B \int \df{r^2}{\Delta} dr,\ \text{with}\ \frac{\pi}{2} - \theta \ll 1,
 \label{eq:EulerKerr}
 \end{equation} 
in terms of the Euler potentials, where $q$ is the monopole charge, 
the angular velocity of the magnetic field line $\Omega_F$ is a free parameter here, and 
$J_B$ is given by the regularity condition of $F_{\mu\nu}F^{\mu\nu}$ at the horizon as
    \begin{equation}
        J_B=\frac{r_\text{H}^2+a^2}{r_\text{H}^2}(\Omega_\text{H}-\Omega_F).
    \label{eq:znajek}
    \end{equation}
 Note that solution \eqref{eq:EulerKerr} is valid for arbitrary values of the spin parameter $a$, but it is consistent with the solution of magnetic field lines for a slowly rotating black hole obtained by \cite{Blandford1977} (see also \cite{Yang2014, Gralla2014}).
 The derivation of solution \eqref{eq:EulerKerr} is discussed in the Appendix.
    
 The physical meaning of the Euler potentials is as follows. The function 
 $\phi_1$ is the so-called stream function, which defines a magnetic 
 surface as $\phi_1=\text{const}$, and $\Omega_{F}$ is, in general, a function of $\phi_1$. 
 Therefore, $\Omega_F$ is a constant for a fixed magnetic surface. 
 The condition $\phi_2=\text{const}$ determines the shape and 
 the time evolution of the magnetic field lines on the magnetic surface. 
 The timelike two-dimensional surface defined by the intersection of $\phi_1=\text{const}$ and 
 $\phi_2=\text{const}$ lying in the four-dimensional spacetime is called the field sheet \cite{Gralla2014}. 
 Considering the above properties, we see that a constant time slice on the field sheet gives the magnetic 
 field line on the magnetic surface at that time.

\begin{figure}[H]
\centering
\includegraphics[width=0.95\linewidth]{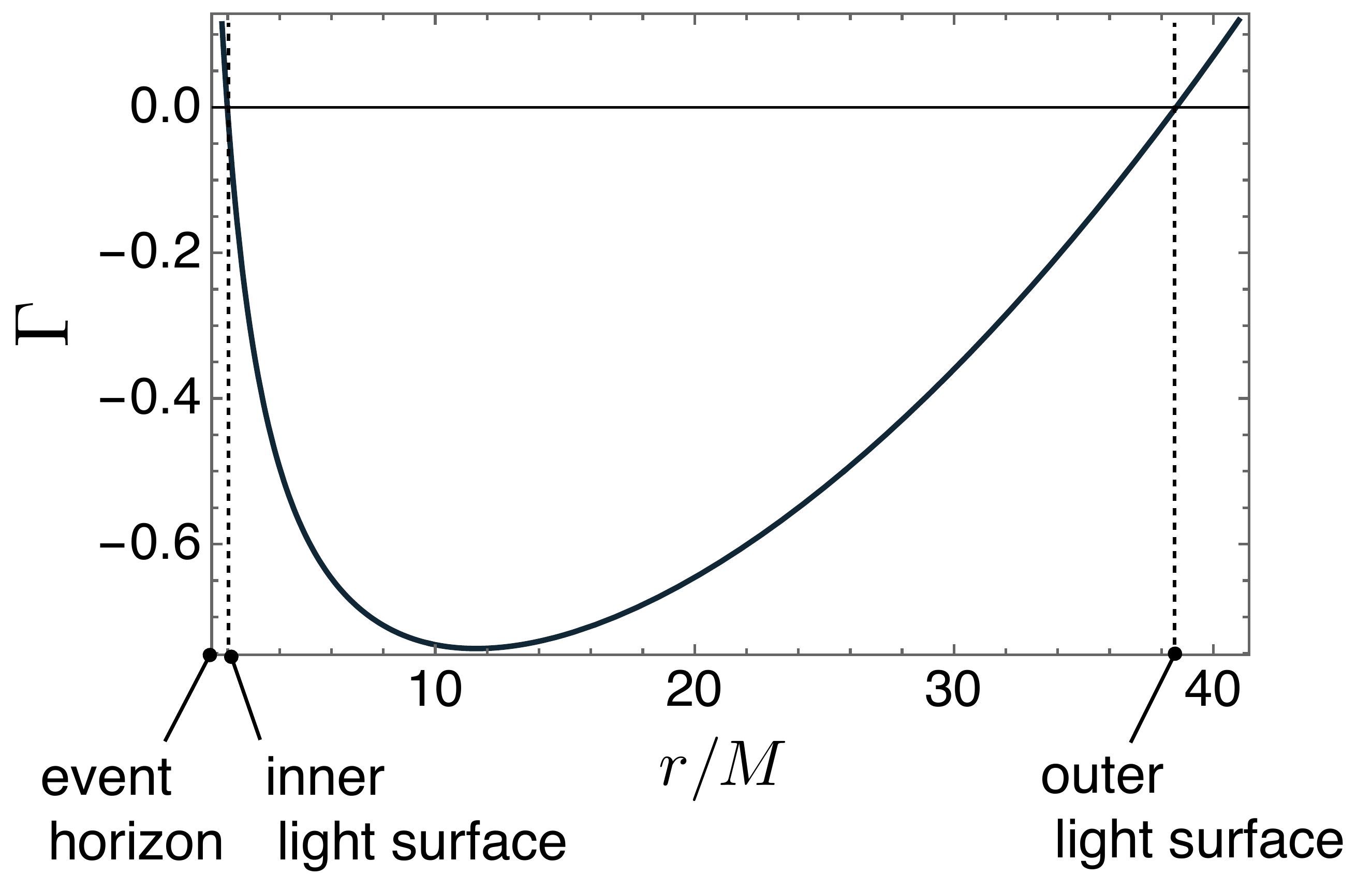}
\caption{\footnotesize{Plot of $\Gamma$ for $a/M=0.2$ and $M\Omega_F=0.027$; $\Gamma<0$ in 
the region between $r_\text{in}$ and $r_\text{out}$, and outside 
the region, $\Gamma$ becomes positive. The left end of this curve is the position 
of the black hole horizon.}}
\label{fig:Gam}
\end{figure}

The background magnetosphere has both the inner and outer light surfaces, which are given as the condition that the 
corotating vector with the magnetic field line $\chi^{\mu}:=\xi_{(t)}^\mu+\Omega_F \xi^{\mu}_{(\vp)}$ becomes null. 
We denote the norm of $\chi^\mu$ by $\Gamma$ and it is evaluated as
\begin{align}
\notag \Gamma&:=
g_{\mu\nu}\chi^{\mu}\chi^{\nu}=g_{tt}+2\Omega_Fg_{t\vp}+\Omega_F^2 g_{\vp \vp}\\
&=-\frac{\Omega_F^2}{r}(r_0-r)(r-r_\text{in})(r-r_\text{out}),\  r_0<0<r_\text{in} <r_\text{out}.
\end{align}
The two positive roots on the equatorial plane $\theta=\pi/2$ are obtained analytically as
\begin{align}
  r_\text{in}&=2d_1\cos\left(\frac{1}{3}\mathrm{arccos}
\left(\frac{d_2}{2d_1}\right)-\frac{2\pi}{3}\right),\\
  r_\text{out}&=2d_1\cos\left(\frac{1}{3}\mathrm{arccos}
\left(\frac{d_2}{2d_1}\right)\right),
\label{eq:rin_rout}
\end{align}
and these are the radii of the inner and outer light surfaces, respectively. The negative root $r_0$ is
\beq
r_0=2d_1\cos\left(\frac{1}{3}\mathrm{arccos}\left(d_2/(2d_1)\right)+2\pi/3\right),
\eeq
where
\beq
 d_1=\left(\frac{1-a^2\Omega_F^2}{3\Omega_F^2}\right)^{1/2},\ \  d_2=-6M\left(\frac{1-a\Omega_F}{1+a\Omega_F}\right).
\eeq
The inner light surface is located outside the black hole horizon: $r_\text{H} < r_\text{in}$, 
which is always satisfied for the present background magnetosphere. 
As shown in Fig.~\ref{fig:Gam}, $\Gamma$ is negative in the region between the light surfaces.

The light surfaces are causal surfaces for propagation of \alf waves \cite{Gralla2014}. 
For all computations in the present paper, we consider \alf wave propagation 
within the range $r_\text{in} \leq r \leq r_\text{out}$ where $\chi^\mu$ is timelike and the velocities of corotating observers are less than 
the speed of light. The outer light surface forms due to the fact that the velocity of the magnetic field lines becomes 
faster and faster at a distant point, then finally, it 
exceeds the speed of light at a far point; whereas, the inner light surface is due to the effect of the gravitational redshift: 
Near the black hole, the speed of light is relatively slow and the velocity of the magnetic field lines becomes larger than the speed of light. 
  \subsection{Energy and angular momentum flux}
  The energy and angular momentum flux vectors are defined with the timelike and spacelike Killing vectors as 
  \beq
  P^\mu=-T^{\mu}_{\ \ \nu}\xi^{\nu}_{(t)},\quad L^\mu=T^{\mu}_{\ \ \nu} \xi^{\nu}_{(\vp)}.
  \label{eq:fluxes}
  \eeq
  Evaluating the radial components of these vectors \cite{Blandford1977}, we obtain
\begin{align}
&P^r=-g^{rr}T_{rt}=-g^{rr}g^{\theta\theta} F_{r\theta}F_{t\theta}\simeq  \Omega_F J_B \frac{q^2}{r^2}\sin^2{\theta},\\
&L^r=g^{rr}T_{r\vp}=g^{rr}g^{\theta\theta} F_{r\theta}F_{\vp \theta}\simeq J_B \frac{q^2}{r^2}\sin^2{\theta},
\label{eq:poyntingflux}
\end{align}
where we consider $\pi/2 -\theta \ll 1$.
As $J_B \propto (\Omega_\text{H}-\Omega_F)$, both the energy and angular momentum fluxes become outward only if
\beq
0 < \Omega_F < \Omega_\text{H}.
\label{eq:BZ2}
\eeq 
This is the condition for occurrence of the BZ process, and if $\Omega_F=\Omega_\text{H}/2$, then the energy flux takes the maximum.

\section{Alfv\'en waves}
In this section, we apply a perturbation to the background magnetosphere obtained in the previous section and discuss the wave propagation in the magnetosphere. There are two different wave modes in the force-free magnetosphere: the fast magnetosonic and Alfv\'en wave, 
which is a longitudinal wave mode due to the magnetic and gas pressure, and a transverse wave mode propagating along magnetic field line due to the magnetic tension. 
In general, these wave modes are coupled to each other, but they can be 
decoupled by considering the perpendicular perturbation to a magnetic surface. 
\subsection{Perturbation and wave modes}
Let $\delta\phi_{i}$ be a perturbation to the Euler potential $\phi_{i}$ for $i=1,2$. 
To define the perturbation, it is useful to introduce the displacement 
vector \cite{Uchida1997c,Uchida1997b} in the $\theta$ direction, whose component is denoted by $\zeta^\theta$. 
Taking the inner product between the derivative of the Euler potentials of the background magnetosphere 
and $\zeta^{\mu}:=\delta^\mu_{\theta}\zeta^{\theta}$, the perturbations 
are obtained as
\beq
\delta \phi_1 =\zeta^{\mu}\partial_{\mu} \phi_1=\zeta^{\theta}(t,r,\vp)\partial_{\theta} \phi_1,\ \ 
\delta \phi_2 =\zeta^{\mu}\partial_{\mu} \phi_2=0
\eeq
Note that we choose the magnetic surface on the equatorial plane of 
the Kerr spacetime; therefore, the first derivative of $\delta \phi_1$ with respect to $\theta$ becomes zero 
due to the definition of the perturbation and the $\theta$ dependence of the background magnetosphere solution \eqref{eq:EulerKerr}.
This indicates that the wave mode $\delta \phi_1$ does not propagate in the $\theta$ direction; specifically, the propagation of this 
wave mode is restricted on the magnetic surface. Moreover, its oscillation is in the perpendicular direction of the magnetic surface; therefore, $\delta \phi_1$ is
 a transverse wave mode propagating on a magnetic surface (\alf wave).
Meanwhile, $\delta \phi_2$ corresponding to the fast magnetosonic wave does not appear for the present perturbation to the background magnetosphere. 
 
\begin{figure}[H]
\centering
\includegraphics[width=0.95\linewidth]{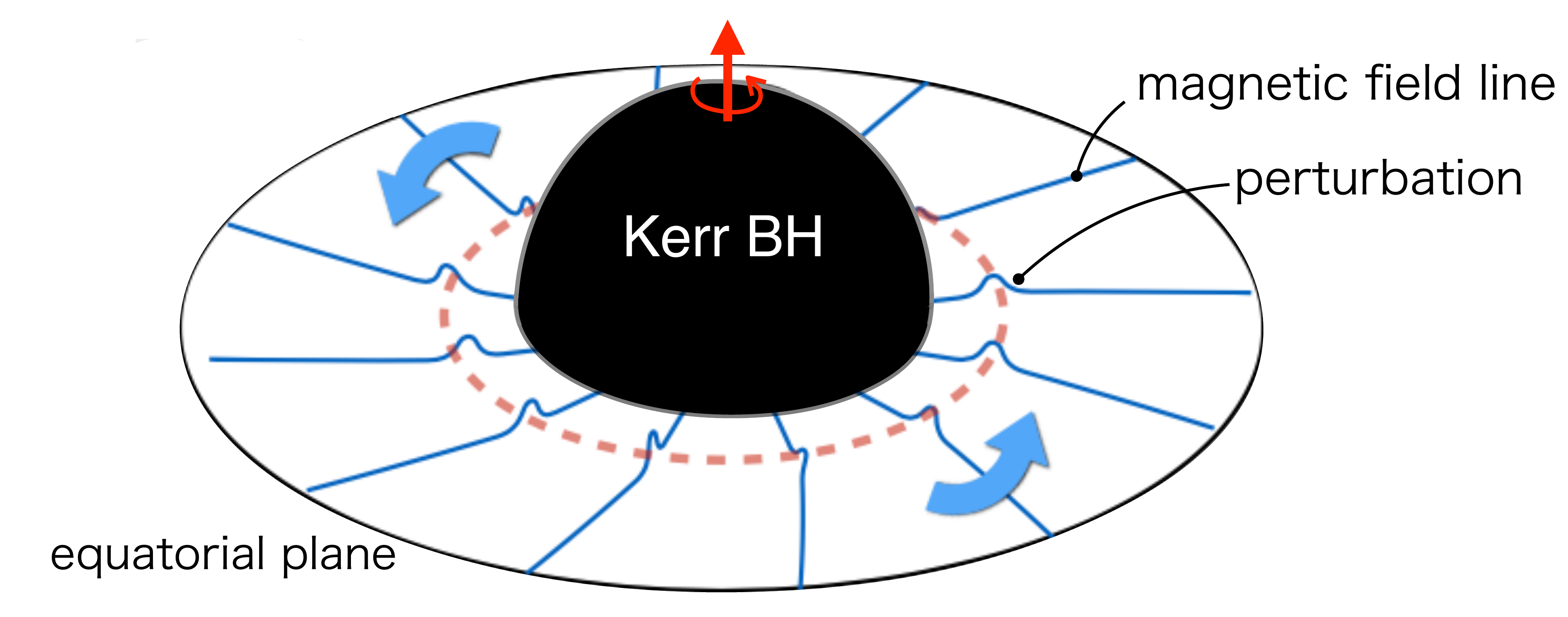}
\caption{\footnotesize{Schematic of the perturbation perpenducular to the magnetic surface on the equatorial plane.}}
\label{fig:cartoon}
\end{figure}

From \eqref{eq:basicEuler}, the first-order perturbation equations are 
\begin{align}
\nonumber &\partial_{\mu}\delta \phi_i \partial_{\nu}\left[\sqrt{-g}\partial^{[\mu}\phi_1\partial^{\nu]} \phi_2 \right]\\
& \quad \quad +\partial_{\mu}\phi_i\partial_{\nu}\left[\sqrt{-g}\left(\partial^{[\mu}\delta \phi_1 \partial^{\nu]}\phi_2 \right)\right]
=0.
\label{eq:waves}
\end{align}
For $i=1$, the second term is zero due to $\pa_\theta \delta \phi_1=0$, whereas the first term is proportional to $\pi/2-\theta\ (\ll 1)$ by expanding this quantity and becomes zero on $\theta=\pi/2$. Therefore, \eqref{eq:waves} with $i=1$ is a trivial equation. For $i=2$, 
we obtain
\beq
\partial_{\mu}\phi_2 \pa_\nu \left[\sqrt{-g}\partial^{[\mu}\delta \phi_1 \partial^{\nu]}\phi_2\right]=0.
 \label{eq:Alfven}
\eeq
This is the wave equation governing the propagation of \alf waves on the magnetic surface at $\theta=\pi/2$. 
A schematic of magnetic fields and the perturbation perpendicular to the magnetic surface ($\theta=\pi/2$) is displayed 
in Fig.~\ref{fig:cartoon}. 
Note that the background magnetic field lines can have curvature in the toroidal direction, which stems from the nonzero $B^\varphi$ \eqref{eq:EB}.
\noindent 

\subsection{\alf wave equation in the form of the Schr\"odinger equation}

In this subsection, we rewrite wave equation \eqref{eq:Alfven} in the form of a Schr\"odinger-type equation by introducing 
a ``tortoise coordinate'', which maps the inner and outer light surfaces to $-\infty$ and $+\infty$, respectively.

In terms of 
the Euler potentials, a magnetic field line and its time evolution is given by 
\beq
\phi_2=\varphi-\Omega_F t  + J_B \int \df{r^2}{\Delta} dr =\text{const.}
\eeq
Considering the property of \alf waves that propagate along a magnetic field line,
we should assume the dependence of variables as 
\beq
\delta \phi_1 =  \delta \phi_1 \left(t,\ r,\ \varphi-\Omega_F t  + J_B \int (r^2/\Delta) dr\right).
\eeq
Substituting this $\delta \phi_1$ into \eqref{eq:Alfven}, we obtain
\begin{align}
\nonumber 
&-\df{r^2}{\Delta}H \pa_t^2  \delta \phi_1+\pa_r\left[-\Gamma\left(\pa_r-\dfrac{{J_B}r^2g_{\varphi \varphi}}{\Gamma \Delta}(\Omega-\Omega_F) \pa_t 
\right) \delta \phi_1 \right]\\
&+\dfrac{{J_B}r^2 g_{\varphi \varphi}}{\Delta}(\Omega-\Omega_F)\pa_t\pa_r  \delta \phi_1
-|\partial \phi_2|^2\  \delta \phi_1=0,
\label{eq:Alfven_sigtau}
\end{align}
where $|\partial \phi_i |^2:=\pa_\mu \phi_i \pa^{\mu}\phi_i$ 
and $H:=1+\Omega_F g_{\vp\vp} \pa_r \phi_2$. 
Note that $|\pa \phi_2|^2$ is proportional to the absolute square of the field strength as 
shown below:
\beq
 \df{F_{\mu\nu}F^{\mu\nu}}{2}=\left[ |\pa \phi_1|^2 |\pa \phi_2|^2 -(\pa^\mu \phi_1\pa_\mu \phi_2)^2 \right]=\df{q^2}{r^2}|\pa \phi_2|^2.
  \label{eq:Fandphi2}
\eeq

\begin{figure}[H]
\centering
\includegraphics[width=1\linewidth]{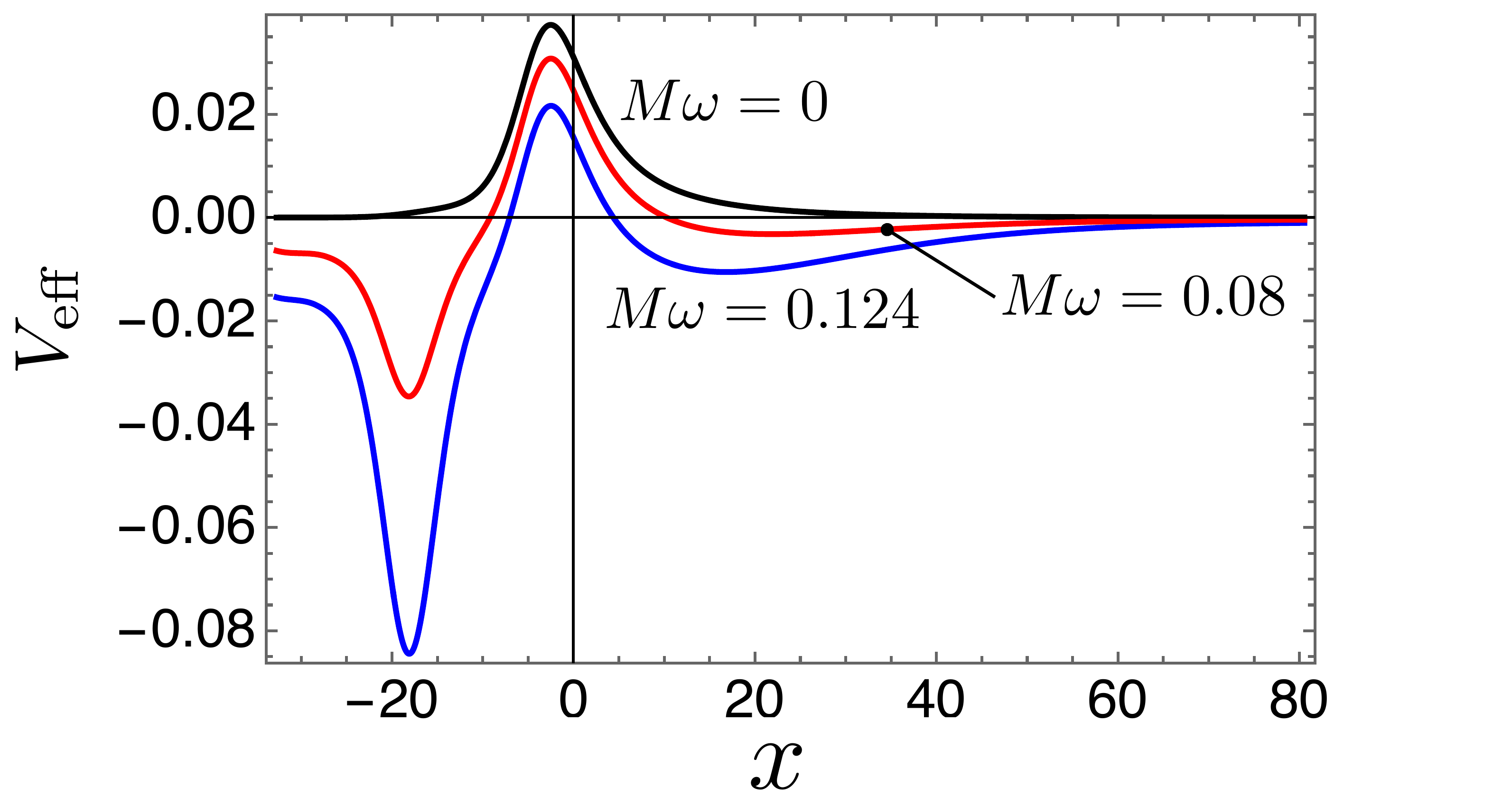}
\caption{\footnotesize{Effective potentials $V_\text{eff}$ for \alf waves propagating on the 
magnetic surface in the vicinity of the equatorial plane for  
$a/M=0.2$, $M\Omega_F=0.041$ with 
$M\omega=0, 0.08, 0.124$. The parameter set $(M\Omega_F, M\omega)=(0.041, 0.124)$ gives a  deep 
bottom, and a resonance can be seen as displayed in Fig.~\ref{fig:plot3d}.}}
\label{fig:veff}
\end{figure}

First, we introduce the following new
coordinates to eliminate the $ \pa_t \pa_r  \delta \phi_1$ term in \eqref{eq:Alfven_sigtau}:
\begin{equation}
\partial_T =\partial_t,\quad\partial_X=\partial_r -\dfrac{J_B r^2g_{\vp \vp}}{\Gamma\Delta}(\Omega-\Omega_F)\partial_t.
\label{eq:newcoordinates}
\end{equation}
The relation between the old and new coordinates is
\begin{equation}
r=X,\quad  t=T- J_B \int dX  \dfrac{X^2g_{\vp \vp}}{\Gamma \Delta}(\Omega-\Omega_F),
\label{eq:rX_tT}
\end{equation}
and equation \eqref{eq:Alfven_sigtau} yields
\begin{align}
\notag &\left[  -\frac{X^2}{\Delta}H +\frac{J_B^2 X^4 g_{\vp\vp}^2}{\Gamma \Delta^2}(\Omega-\Omega_F)^2\right]\pa_T^2  \delta \phi_1\\
& \quad \quad +\partial_X \left(-\Gamma \partial_X  \delta \phi_1 \right)-|\partial \phi_2|^2  \delta \phi_1 =0.
\label{eq:waveeq_TX}
\end{align}
Then, separating the variables as $ \delta \phi_1=e^{-i\omega T}R(X) $ and 
introducing the ``tortoise'' coordinate $x$ as $dx/dX=-\Gamma^{-1}$, we obtain the Schr\"{o}dinger-type equation\footnote{The inner light surface is 
a causal boundary like a black hole horizon for \alf waves. Therefore, it is useful to map the point to $-\infty$ as in the case of the 
analysis of the black hole perturbation equation.}
\begin{align}
&\dfrac{d^2 R }{dx^2}-V_\text{eff}R=0,
\label{eq:Sch}  \\
&V_\text{eff}=-\Gamma |\pa \phi_2|^2+\df{\omega^2X^2}{\Delta} \left[ H\Gamma -\frac{ J_B^2 X^2 g_{\vp\vp}^2 }{\Delta}(\Omega-\Omega_F)^2 \right].
\label{eq:Veff}
\end{align}
In the tortoise coordinate, the locations of the inner and outer light surfaces becomes 
$x=-\infty$ and $x=+\infty$, respectively. The effective potentials for several frequencies are plotted in Fig.~\ref{fig:veff}.
\noindent
Note that the $x$-dependence of the terms with $\omega$ in \eqref{eq:Veff} is very small, as seen from Fig.~\ref{fig:veff}. 
Therefore, properties of the effective potential, such as the existence and the position of the peak, are determined by the first term, 
which reflects the effect of the gravitational redshift and the angular velocity of magnetic field lines in $\Gamma$ 
as well as the square of the field strength.

\section{Alfv\'enic superradiance and resonant scattering}
In this section, we discuss the wave scattering problem of \alf waves by solving 
Schr\"{o}dinger-type equation \eqref{eq:Sch} for the corotating coordinates $(T,x)$. 
For \alf waves to be scattered by the effective potential efficiently, we focus only on the 
relatively low frequency 
cases: $M\omega=0.050$ - $0.250$. 
From the coefficients of the asymptotic ingoing and outgoing solutions, we define the 
reflection rate of the \alf waves, then obtain the condition for \alfv superradiance. 

\subsection{Reflection rate of \alf waves and the condition for \alfv superradiance}
To evaluate the asymptotic form of the wave function $R$ in Eq.~\eqref{eq:Sch}, first, 
we examine the asymptotic form of the effective potential. Then, the definition of ingoing 
and outgoing modes in this scattering problem is discussed.  As $\Gamma \simeq 0$ 
in the vicinity of the two light surfaces, the asymptotic form of the effective potential is
\vspace{-0.4cm}
\beq
V_\text{eff}^{\text{asymp}}\simeq -\df{\omega^2 J_B^2  
X^4 g_{\vp\vp}^2}{\Delta^2}(\Omega-\Omega_F)^2 \ <0.
\eeq
Therefore, the asymptotic solution of Eq.~\eqref{eq:Sch} is written in the following form:
\begin{align}
\notag R &\propto \exp{\left[ \pm i \int dx \sqrt{-V_\text{eff}^{\text{asymp}}} \right]} \\
&=\exp{\left[ \pm i  \omega \int dx \df{X^2 g_{\vp\vp}}{\Delta }|J_B|(\Omega_F-\Omega)\ \right]},
\label{eq:Rinout}
\end{align}
where $\omega, X^2, g_{\vp\vp}$, and $\Delta$ are positive definite quantities, whereas 
the sign of $\Omega-\Omega_F$ can be changed depending on the value 
of $\Omega_F$ and the location $r$. At a point far from the black hole where the dragging effect of the spacetime is almost zero: 
$\Omega \sim 0$ and the sign of the integrand in \eqref{eq:Rinout} is positive. Therefore, 
the positive (negative) sign in \eqref{eq:Rinout} indicates the outgoing (ingoing) wave there. 
We use this asymptotic behavior of the phase of the wave function to define the in and outgoing modes.

\begin{figure}[H]
\centering
\includegraphics[width=0.75\linewidth]{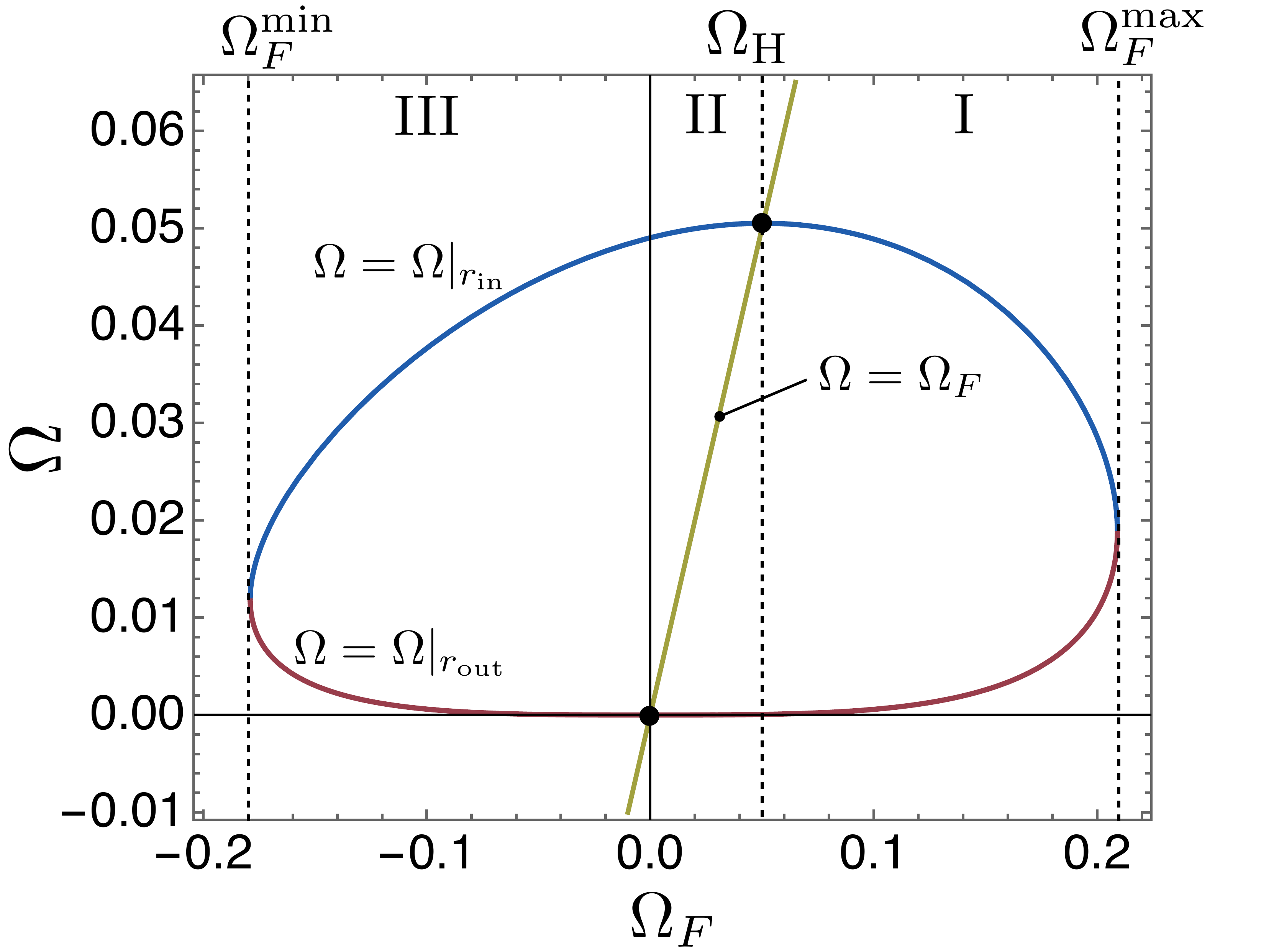}
\caption{\footnotesize{Relationship among functions $\Omega=\Omega|_{r_\text{in}}(\Omega_F), \Omega=\Omega|_{r_\text{out}}(\Omega_F)$, and $\Omega=\Omega_F$ for a fixed spin parameter $a/M=0.2$. 
The black dots correspond to the solutions of $\Omega|_{r_\text{in}}(\Omega_F)=\Omega_F$ and 
$\Omega|_{r_\text{out}}(\Omega_F)=\Omega_F$, which are $\Omega_F=\Omega_\text{H}$ and $\Omega_F=0$, respectively.
The minimum and maximum of $M\Omega_F$ for the existence of two light surfaces are denoted by $\Omega_F^\text{min}$ and $\Omega_F^\text{max}$, respectively.}}
\label{fig:omLS}
\end{figure}

As the inner light surface is the causal boundary for \alf waves \cite{Gralla2014}, we require the purely 
ingoing boundary condition at the inner light surface.
The asymptotic solutions of Eq.~\eqref{eq:Sch} with the 
ingoing boundary condition at the light surfaces are
\begin{widetext}
\begin{equation}
R=
\begin{cases}
{\displaystyle\exp{\left[ - i  \omega \int \frac{dx}{\Delta} X^2g_{\vp\vp}|J_B|(\Omega_F-\Omega) \right]}}
\quad\quad \quad\quad\quad\quad\quad\quad\quad\quad\quad\quad\quad\quad\quad 
\quad\quad\quad \quad \ \text{for}\quad  x \rightarrow -\infty,\\
{\displaystyle 
 A_\text{in}{\displaystyle\exp{\left[ - i \omega \int \frac{dx}{\Delta} X^2g_{\vp\vp}|J_B|(\Omega_F-\Omega) \right]}}
+A_\text{out}{\displaystyle\exp{\left[  i\omega \int \frac{dx}{\Delta} X^2g_{\vp\vp}|J_B|(\Omega_F-\Omega) \right]}}} \ \text{for}\quad  x\rightarrow+\infty.
\end{cases}
\label{eq:asympt1}
\end{equation}
\end{widetext}
The ingoing wave around the inner light surface becomes outward when the spacetime dragging effect is so large that the sign 
of the integrand get flipped\footnote{This point is similar to superradiance for other waves such as scalar waves.}.
From the conservation of the Wronskian, we obtain the reflection rate
of the wave as
\begin{equation}
\left|\frac{A_{\text{out}}}{A_{\text{in}}}\right|^{2}=1-\dfrac{f_\text{in}}{f_\text{out}}\dfrac{(\Omega_F - \Omega|_{r_\text{in}})}{(\Omega_F- \Omega|_{r_\text{out}})}\dfrac{1}{|A_\text{in}|^2}\ ,
\label{eq:ref}
\end{equation}
where $f_\text{in/out}=(X^2g_{\vp\vp}/\Delta)|_{r_\text{in/out}}$. 
From Eq.~\eqref{eq:ref}, one sees that the reflection rate $|A_\text{out}/A_\text{in}|^2$ exceeds 
unity and the reflected \alf wave will be amplified through the scattering by the effective potential 
(\alfv superradiance) if the angular velocity of the magnetic field line satisfies
\footnote{In BTZ black string case \cite{Noda2020}, 
the superradiant condition is $0<\Omega_F <\Omega|_{r_\text{in}}$ because there is not outer light surface due to the asymptotic AdS structure of the spacetime.}
\begin{equation}
\Omega|_{r_\text{out}}<\Omega_F<\ \Omega|_{r_\text{in}}.
\label{eq:alf_superrad}
\end{equation}
Note that the functions $\Omega|_{r_\text{in/out}}$ depend on $\Omega_F$, hence 
we need to solve the inequality for $\Omega_F$ to evaluate it. As the functions 
$\Omega|_{r_\text{in/out}}$ are too algebraically complex to solve, instead of that, we plot those functions of $\Omega_F$ in Fig.~\ref{fig:omLS}. The superradiant condition \eqref{eq:alf_superrad} holds only in the region II where 
\beq
0<\Omega_F<\Omega_\text{H},
\label{eq:BZ3}
\eeq
 in Fig.~\ref{fig:omLS}. 
Therefore, the superradiant condition \eqref{eq:alf_superrad} is exactly the same as the condition for the BZ process \eqref{eq:BZ2}.
\begin{figure}[H]
\centering
\includegraphics[width=1\linewidth]{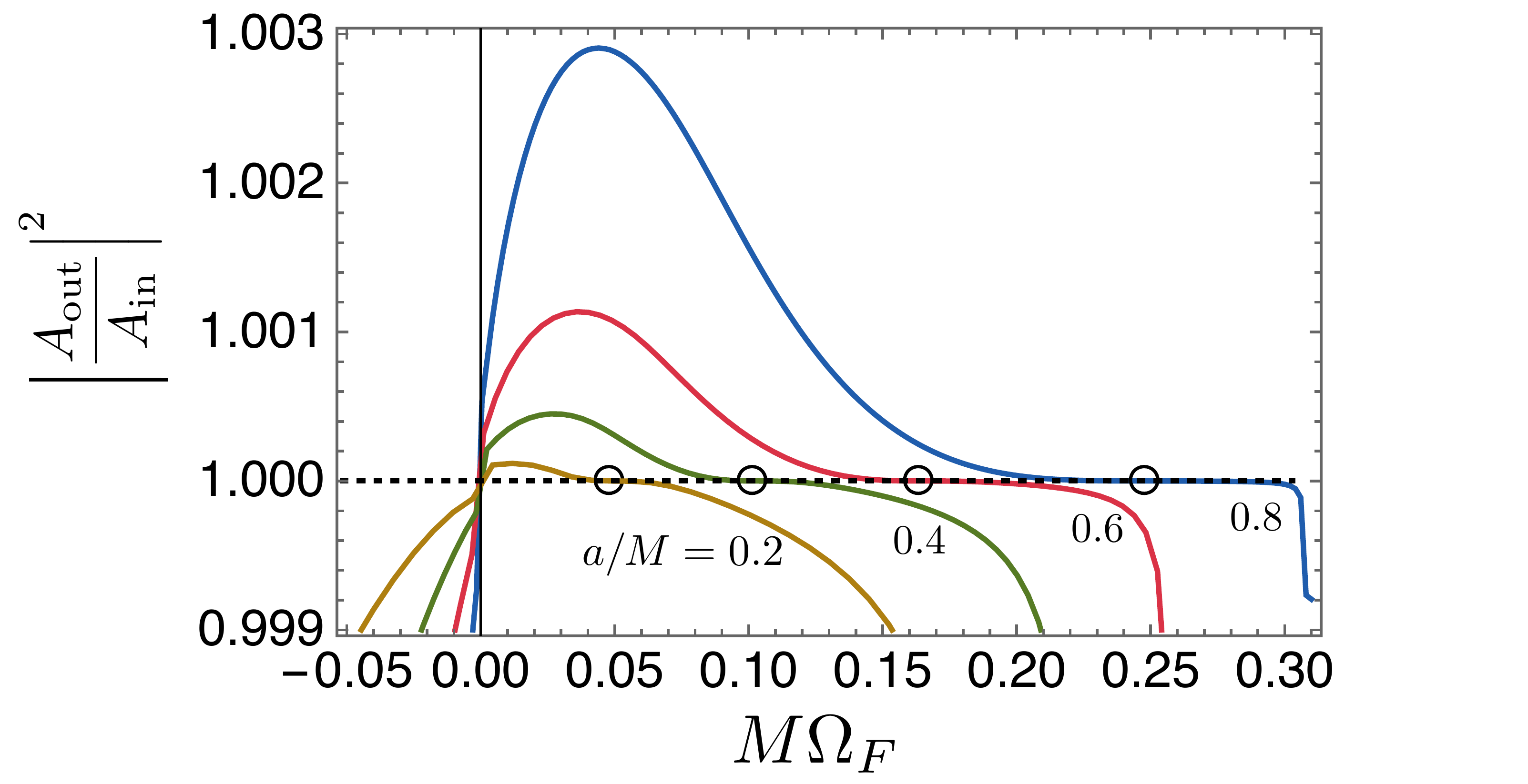}
\caption{\footnotesize{Reflection rate for the $M\omega=0.02$ case with several spin parameters. The threshold for $M\Omega_F$, which is marked as $\circ$, is given by the angular velocity of the black hole horizon, as we expected from 
Eq.~\eqref{eq:alf_superrad},  Eq.~\eqref{eq:BZ2}, and Fig.~\ref{fig:omLS}. The values are $M \Omega_\text{H}=0.0505,0.104,0.167,0.250$ for $a/M=0.2, 0.4,0.6,0.8$, respectively.}}
\label{fig:reflection_rate}
\end{figure}
%
\begin{figure}[H]
\centering
\includegraphics[width=0.85\linewidth]{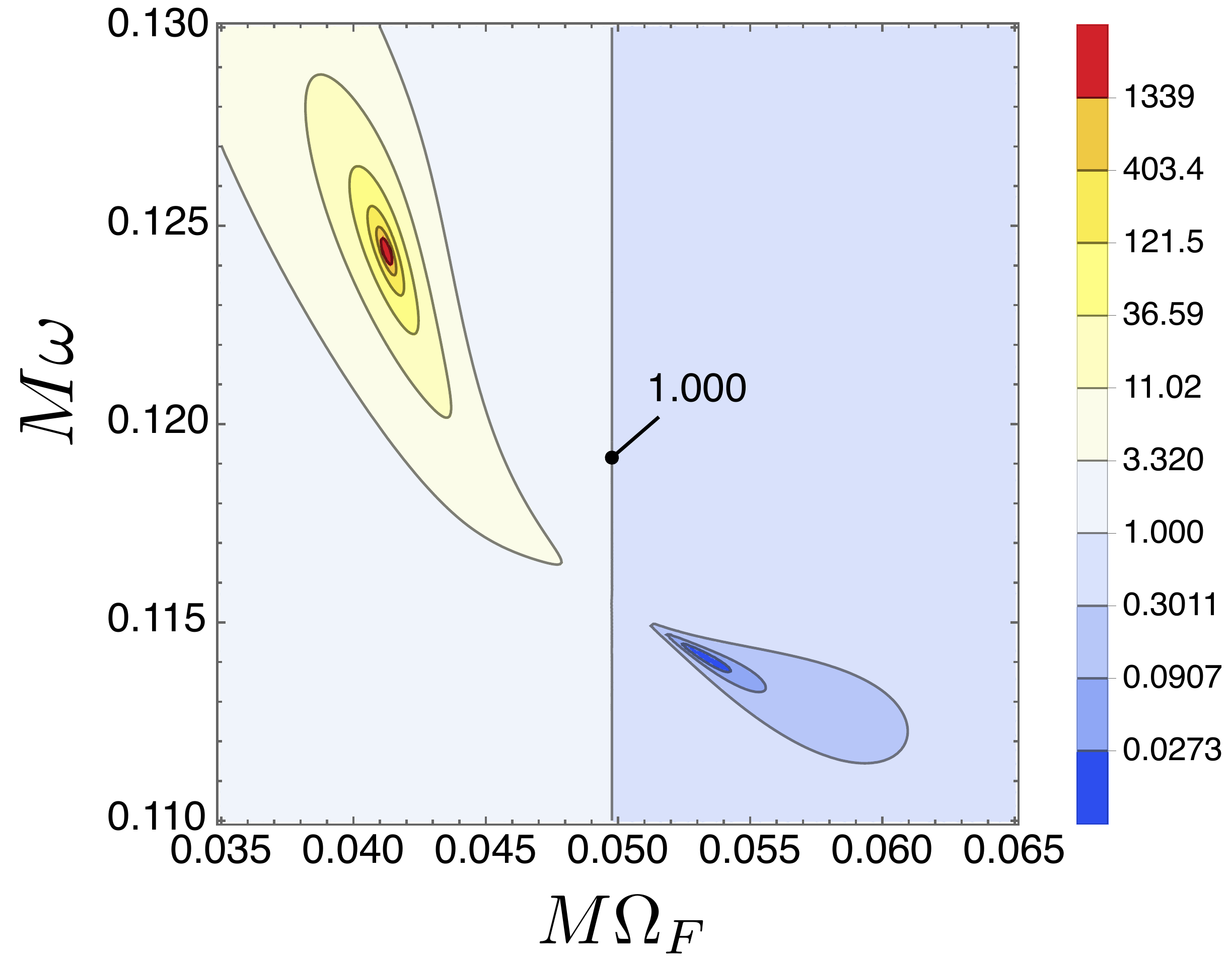}
\caption{\footnotesize{Contour plot of $|A_\text{out}/A_\text{in}|^2$ on $\Omega_F$-$\omega$ plane for the $a/M=0.2$ case. The vertical line labeled as $1.000$ is the contour of $|A_\text{out}/A_\text{in}|^2=1$.}}
\label{fig:plot3d}
\end{figure}
%
\begin{figure}[H]
\centering
\includegraphics[width=1.\linewidth]{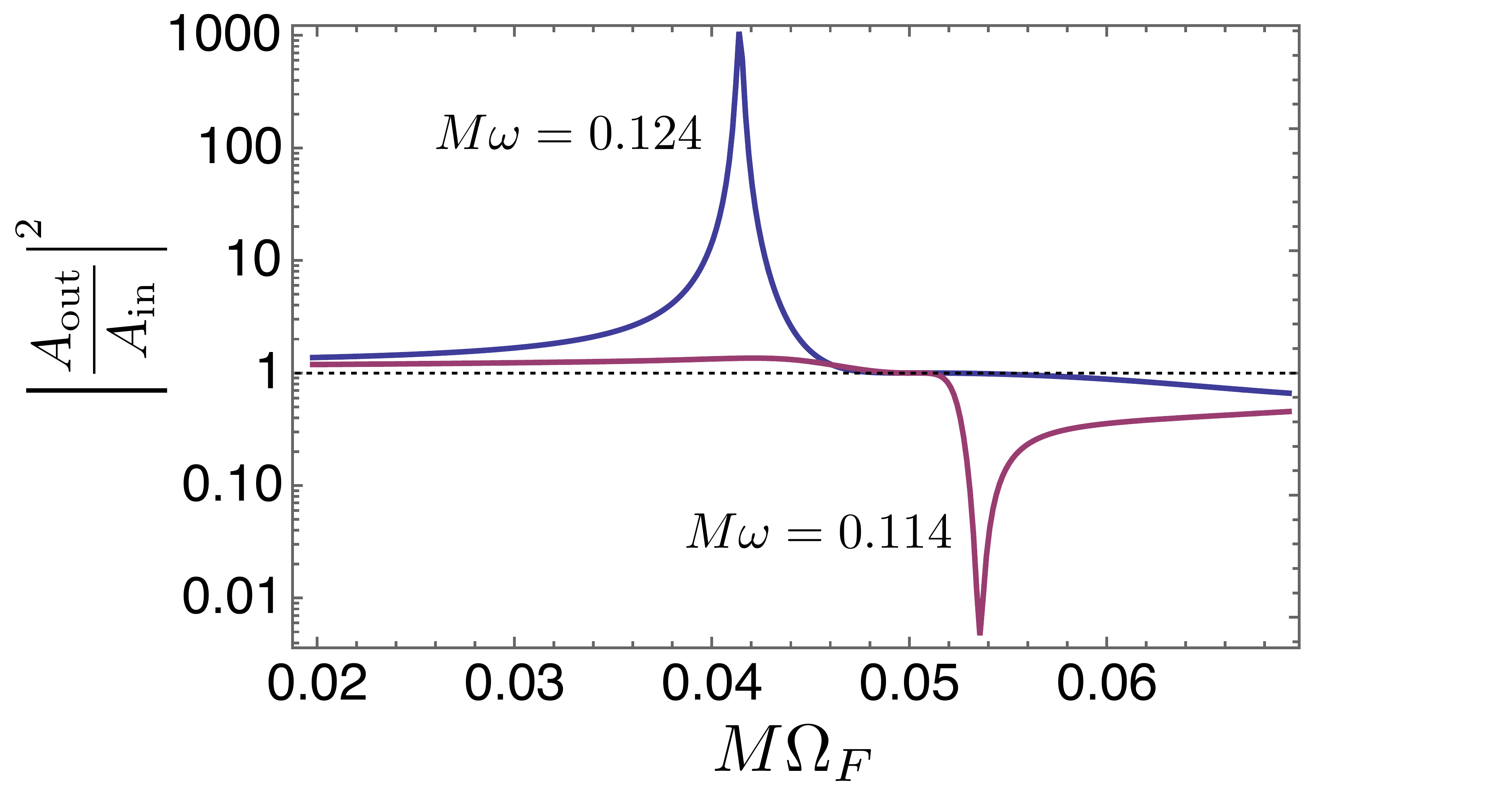}
\caption{\footnotesize{Reflection rates for the frequencies giving the resonant scattering and absorption in the $a/M=0.2$ case.}}
\label{fig:refrate_resonance}
\end{figure}
 \noindent 
 For $M \Omega_F$ in regions I--III, we evaluate the reflection rate by solving wave equation \eqref{eq:Sch} numerically. 
 The results for various spin parameters of the Kerr spacetime $a/M$ with 
fixed $M\omega$ are shown in Fig.~\ref{fig:reflection_rate}.  
Indeed, the reflection rate exceeds unity for 
$M \Omega_F$ satisfying superradiant condition \eqref{eq:alf_superrad} or equivalently \eqref{eq:BZ2}.  

In Fig.~\ref{fig:plot3d}, we present the contour plot of the reflection rate on the $\Omega_F$-$\omega$ plane for $a/M=0.2$. 
The contours of $|A_\text{out}/A_\text{in}|^2=1$ correspond to $M\Omega_F=0$ and $M\Omega_F=M\Omega_\text{H}$. \alfv superradiance occurs 
in the region between these two lines. There is a peak associated with $A_\text{in}\sim 0$ at $(M\Omega_F, M\omega)=(0.0412, 0.124)$. 
As this situation is similar to the quasinormal modes of black hole perturbation, which come from the boundary condition $A_\text{in}=0$ with complex frequency, we search the frequency in the complex plane of $\omega$ with the fixed angular velocity of the magnetic field line $M\Omega_F=0.0412$. As a result, we realized a frequency giving $A_\text{in}=0$ at $M\omega = 0.1240- 0.0002 \ i$.
 Therefore, the peak in Fig.~\ref{fig:plot3d} reflects a resonant scattering corresponding to a ``quasinormal mode" of the magnetosphere for \alfv perturbation\footnote{The quasinormal modes of magnetosphere itself have already been discussed in \cite{Yang2014} although it is not for \alfv perturbation.}. As the imaginary part of those frequencies are negative, the 
present magnetosphere is stable for the perturbation with \alfv superradiance. 
Moreover, there is a bottom at $(M\Omega_F, M\omega)=(0.054, 0.114)$. The presence of the bottom comes from $A_\text{out}=0$, which corresponds to a resonant absorption of \alf waves. 
For those two cases with resonant scattering, the reflection rates are plotted in Fig.~\ref{fig:refrate_resonance}.


\section{Discussion and Concluding Remarks}


In this paper, based on the force-free approximation, we discussed Alfv\'enic superradiance 
in the Kerr spacetime to investigate the difference from our previous work for the BTZ black string spacetime \cite{Noda2020}. 
The structure of the background magnetic field lines considered here is a monopole-like in the poloidal plane, and the inner and outer 
light surfaces exist. We investigated the propagation of Alfv\'en waves by applying a perturbation 
perpendicular to the magnetic surface in the vicinity of the equatorial plane of the Kerr spacetime.

Introducing the tortoise coordinate $x$, the wave equation for \alf waves can be written in the form of the Schr\"{o}dinger-type equation. 
To investigate the reflection rate, we defined the in and outgoing waves at asymptotic regions near the inner and outer light surfaces. 
Then, considering the conservation of the Wronskian, we derived the superradiant condition for \alf waves, which is exactly the same as 
that for the BZ process. Due to the existence of the outer light surface, the superradiant condition in the Kerr spacetime appears to be 
slightly modified from the condition derived in \cite{Noda2020}; however, in Fig.~\ref{fig:omLS}, both are shown to be the same as 
the condition for the BZ process after all. 

The result of this study demonstrates that \alfv superradiance, which was discussed only for the magnetosphere around a BTZ black string \cite{Noda2020}, is possible for the Kerr spacetime case as well. Therefore, it would be important for the extraction process of the rotational energy 
of astrophysical black holes regarding relativistic jets and/or high-energy radiations in active galactic nuclei or gamma ray bursts. 
In particular, the resonant scattering is determined by not only the frequency of \alf waves, but also 
the parameter of the mangetosphere such as 
$\Omega_F$, and the structure of magnetosphere, specifically that it provides the shape of the effective potential. If we observe 
this resonant scattering as a burst-like emission of electromagnetic waves, information on the structure of the 
magnetosphere and the black hole spacetime would be derived. 

The dynamical situation and higher order of the perturbation are also important, as discussed in the recent work \cite{Koide2021}. 
A higher order of perturbation can generate a richer phenomenon, as suggested in \cite{Koide2021}: The second order perturbation to $\phi_2$ obeys the Klein-Gordon equation with a source term determined by $\delta \phi_1$. 
Specifically, the linear \alf waves can evoke the second order fast 
magnetosonic wave. Regarding this, a nonlinear effect that results in the conversion of \alf waves to fast magnetosonic waves in rotating 
magnetospheres around neutron stars has been discussed in \cite{Yuan2021}.

We restricted the discussion herein to a stationary magnetosphere filled with 
a strong magnetic field, for which the force-free approximation is valid. 
To grasp what really happens around astrophysical black holes, it is necessary to consider the plasma effects and the environment around a black hole such as an accretion disk, and to discuss how the rotational energy extracted by Alfv\'en waves can be transported and converted into the kinetic energy of plasmas and how they contribute to the relativistic jets. We leave these tasks for our next papers.

\vspace{-0.5cm}
\begin{acknowledgments}
  The authors thank Shinji Koide, Hirotaka Yoshino, Kenji Toma, and Hideki Ishihara for fruitful discussions. 
    Y.N. was supported in part by JSPS KAKENHI Grant No. 19K03866. 
    M.T. was supported in part by JSPS KAKENHI Grant No. 17K05439. 
    S.N. gratefully acknowledges the hospitality of Kogakuin University, where this work was partially done.
\end{acknowledgments}

\begin{appendix}
\section{Derivation of the background magnetosphere near the equatorial plane}
Here, we demonstrate the derivation of background magnetosphere solution \eqref{eq:EulerKerr} by solving the 
following basic equation of the force-free electrodynamics:
 \begin{equation}
  \partial_{\mu}\phi_i \partial_{\nu}\left[\sqrt{-g}\left(\partial^{\mu}\phi_1\partial^{\nu}\phi_2-\partial^{\nu}\phi_1\partial^{\mu}\phi_2\right)\right]=0.\ \ \ \ (i=1, 2)
        \label{eq:basicEuler_app}
 \end{equation} 
In general, the Euler potentials for stationary and axisymmetric magnetosphere can be written as 
\begin{equation}
    \phi_{1}=\Psi(r, \theta),\
    \phi_{2}=\varphi-\Omega_{F}(\Psi)t+\Phi(r, \theta).
\end{equation}
This was discussed by Uchida \cite{Uchida1997a} with Killing vectors.
Here, to obtain a force-free magnetosphere in the vicinity of the equatorial plane, we assume that 
functions $\Psi$ and $\Phi$ in the Euler potentials depend on the variables as 
\begin{equation}
    \phi_{1}=\Psi(\theta),\
    \phi_{2}=\varphi-\Omega_{F}t+\Phi(r),
\label{eq:ansatz}
\end{equation}
  where $\Omega_F$ is a constant corresponding to the angular velocity of magnetic field lines.
Substituting this ansatz into Eq. \eqref{eq:basicEuler_app} and expanding it up 
to the first order of the small angle measured from the equatorial plane, we obtain:
\beq
\epsilon=\frac{\pi}{2}-\theta.
\eeq

First, for $i=1$, \eqref{eq:basicEuler_app} yields
\begin{align}
\notag 0&=\pa_\theta \Psi\  \pa_\nu \left( \sin\theta \ \pa_\theta \Psi \pa^\nu \phi_2 \right)\\ 
             &=\sin\theta (\pa_\theta \Psi)^2\ \pa_\nu \left(\pa^\nu \phi_2 \right).
\end{align}
Assuming $\sin\theta (\pa \Psi)^2 \neq 0$, we obtain 
\beq
\pa_r \left( \df{\Delta}{\Sigma}  \pa_r \Phi \right)=0.
\eeq
Considering the fact $\Sigma = r^2+a^2\cos^2{\theta} = r^2 +{\cal{O}}(\epsilon^2)$, the above equation becomes the differential equation 
only for $r$. Then, the solution is 
\beq
\Phi(r) = J_B \int \df{r^2}{\Delta}\ dr.
\eeq
The constant $J_B$ stems from $\Delta/r^2 \pa_r \Phi = \text{const}:=J_B$ and is determined by the regularity of $F_{\mu\nu}F^{\mu\nu}$ at 
the black hole horizon. For $i=2$, \eqref{eq:basicEuler_app} gives
\begin{align}
\notag 0&=\pa_\nu \phi_2 \pa_\theta \left(  \sqrt{-g} \pa^\theta \phi_1 \pa^\nu \phi_2  \right)\\ 
             &=|\pa \phi_2|^2  \pa_\theta \left(\sin \theta\  \pa_\theta \Psi  \right) + \sin \theta\  \pa_\theta \Psi \ \pa_\theta |\pa \phi_2|^2,
             \label{eq:eqforPsi}
\end{align}
where $|\pa \phi_2|^2:=\pa_\nu \phi_2 \pa^\nu \phi_2$, which is expanded as
\beq
|\pa \phi_2|^2= (\text{function of} \ r) + {\cal{O}}(\epsilon^2).
\eeq
The derivative of $|\pa \phi_2|^2$ with respect to $\theta$ is proportional to $\cos \theta$, hence it is ignored in the present approximation.
Therefore, Eq.~\eqref{eq:eqforPsi} finally yields
\beq
\pa_\theta \left(\sin\theta \Psi(\theta)\right)=0,
\eeq
for which, the solution is 
\beq
\Psi(\theta)= q \cos \theta,
\eeq
in the vicinity of the equatorial plane. Here, $q$ represents the monopole charge. 
Thus, the background force-free magnetosphere solution is obtained as \eqref{eq:EulerKerr}.

To investigate the structure of the magnetic field lines, we compute the electro and magnetic fields on the equatorial plane measured by 
a Killing observer whose four velocity is $u^{\nu}=(1,0,0,0)$.  The nonzero components of the electric and magnetic fields are
\begin{align}
&E^\theta= \frac{q\Omega_F}{r^2},\ \ B^r= \df{q}{r^2}(g_{tt}+\Omega_F g_{t\vp}),\\
&B^\varphi= -\df{qJ_B}{r^2}g_{tt}\ \ B^t= \df{q J_B}{\Delta}g_{t\vp}.
\label{eq:EB}
\end{align}
Note that 
this background solution corresponds to a monopole-like magnetosphere in the vicinity of the equatorial plane of the Kerr spacetime.

\end{appendix}
%

\end{document}